\documentclass{anstrans}

\disclaimer{This paper is an original manuscript describing the work presented at the \emph{2$^\text{nd}$ IAEA Technical Meeting on Fusion Data Processing, Validation and Analysis} that took place in Boston, USA, from May 30 to June 2, 2017.}\addtocounter{footnote}{1}

\title{Full-pulse Tomographic Reconstruction with Deep Neural Networks}
\author{Diogo R.~Ferreira,$^{*}$ Pedro J.~Carvalho,$^{*}$ Hor\'{a}cio Fernandes,$^{*}$ and JET Contributors\footnotemark}

\institute{EUROfusion Consortium, JET, Culham Science Centre, Abingdon, OX14 3DB, UK\and	
	$^{*}$Instituto de Plasmas e Fus\~{a}o Nuclear, Instituto Superior T\'{e}cnico, Universidade de Lisboa, 1049-001 Lisboa, Portugal}

\email{diogo.ferreira@tecnico.ulisboa.pt}

\usepackage{url}
\usepackage{graphicx}
\usepackage{booktabs}
\usepackage{microtype}

\usepackage{dblfloatfix}
\usepackage[keeplastbox]{flushend}

\begin{document}

\abstract{Plasma tomography consists in reconstructing the 2D radiation profile in a poloidal cross-section of a fusion device, based on line-integrated measurements along several lines of sight. The reconstruction process is computationally intensive and, in practice, only a few reconstructions are usually computed per pulse. In this work, we trained a deep neural network based on a large collection of sample tomograms that have been produced at JET over several years. Once trained, the network is able to reproduce those results with high accuracy. More importantly, it can compute all the tomographic reconstructions for a given pulse in just a few seconds. This makes it possible to visualize several phenomena -- such as plasma heating, disruptions and impurity transport -- over the course of a discharge.}

\footnotetext{See the author list of ``Overview of the JET results in support to ITER'' by X. Litaudon et al. in \textit{Nuclear Fusion}, \textbf{57}, \textit{10}, 102001 (2017).}

\section{Introduction}
\label{sec:intro}

One way to measure plasma radiation is through the use of bolometers, in particular foil bolometers~\cite{ingesson08tomography}. These bolometers consists of a thin metal foil (about 10~$\mu$m) coupled with a temperature-sensitive resistance. As the metal foil absorbs radiation power, its temperature changes and there is a proportional change in resistance. This can be measured using a standard setup, such as a Wheatstone bridge. Overall, such system provides a linear response to the absorbed power, in the range from ultraviolet (UV) to soft X-ray~\cite{mast91bolometer}.

\begin{figure}[h]
	\centering
	\includegraphics[width=\linewidth]{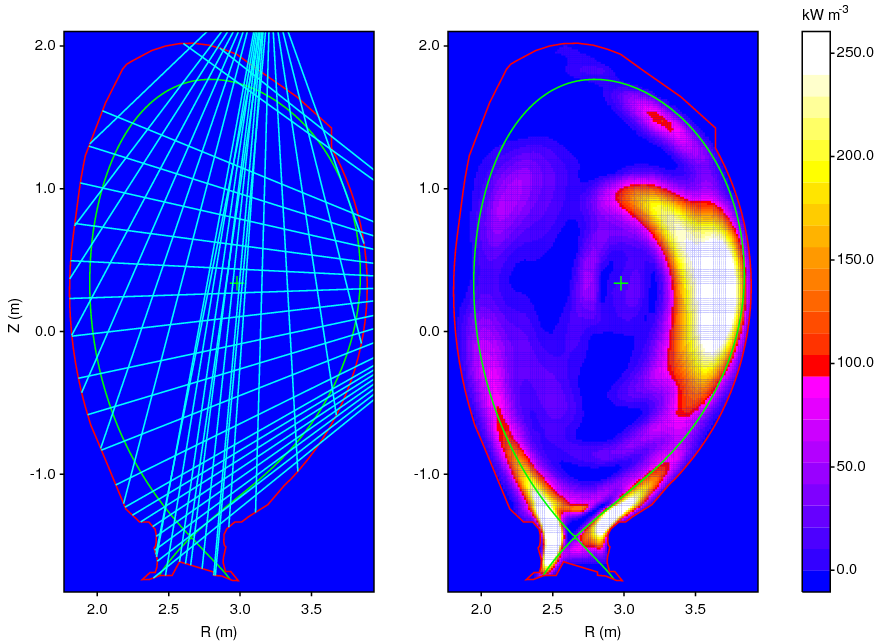}
	\caption{Lines of sight for the vertical and horizontal cameras (\emph{left}) and a sample tomographic reconstruction for pulse 89065 at t=47.8s (\emph{right})}
	\label{fig:lines_tomo}
\end{figure}

At the Joint European Torus (JET) there is a multi-channel bolometer system comprising a horizontal camera and a vertical camera, with 24 bolometers each~\cite{huber07upgraded}. The horizontal camera has a pinhole structure that defines the lines of sight for each of its 24 bolometers. The vertical camera, on the other hand, uses a collimator block to achieve the same purpose~\cite{mccormick05bolometry}. In addition, the vertical camera has an extra 8 bolometers that can be used as reserve channels, so in total the system can provide 56 lines of sight over the plasma, as illustrated in Figure~\ref{fig:lines_tomo}.

This two-camera bolometer system is the basis for many tomographic reconstructions that are routinely performed at JET during post-pulse analysis. The method that is used to compute such reconstructions has been developed by Ingesson et al.~\cite{ingesson98tomography} and actually predates the current bolometer system, having been used with previous generations of soft X-ray diagnostics at JET. In essence, it is an iterative constrained optimization method that minimizes the error with respect to the observed measurements, while requiring the solution to be non-negative. To do so, it solves a generalized eigenvalue problem in order to find a solution as a function of Lagrange multipliers, and then adjusts these Lagrange multipliers iteratively until the non-negativity constraints are satisfied~\cite{fehmers98algorithm}.

Both the solution to the generalized eigenvalue problem (which can be computed using a standard numerical library) and the iterative adjustments to the Lagrange multipliers take a significant amount of computation time. The total run-time depends on the actual data but, with the code available at JET, it can take more than 1h to produce a reconstruction. This makes it impractical to compute more than a few reconstructions per pulse. There is hardly an opportunity to see how the radiation profile develops across an entire pulse.

To appreciate the computational effort involved, consider the following: the bolometer system at JET has a sampling rate of 5~kHz; for the purpose of noise reduction, a window average of 5~ms is usually applied, which corresponds to 25 samples; subsampling by a factor of 25 yields an effective sampling rate of 200~Hz; so, in principle, it should be possible to have as much as 200 reconstructions per second of pulse time; for a pulse of about 30 seconds, this means a total of 6000 reconstructions; at an average of 1h per reconstruction, this would require 250 days.

\begin{figure*}[b]
	\centering
	\includegraphics[width=\textwidth]{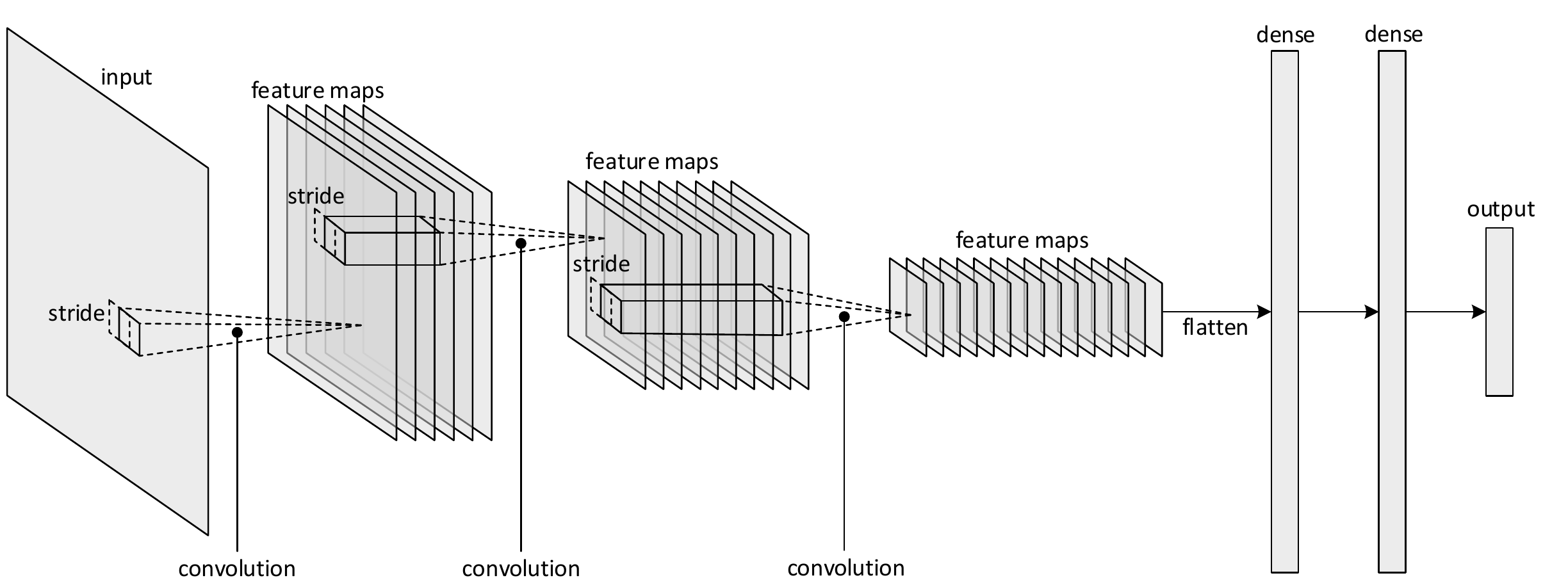}
	\caption{Typical structure of a convolutional neural network (CNN)}
	\label{fig:cnn}
\end{figure*}

Clearly, another way to compute the reconstructions for an entire pulse should be found. The following sections describe how a deep neural network was devised and trained on existing tomograms to produce the same results. A small amount of error is more than compensated by a large computational speedup (both to be quantified below). With this approach, it becomes possible to analyze the time evolution of the radiation profile in great detail. The pulses discussed at the end of this paper, which illustrate disruptions and impurity transport, are only a few examples of what can be done.

\section{Deep neural networks}

Deep learning~\cite{lecun15deep,goodfellow16deep} is having a tremendous impact in fields such as image processing and natural language processing. In particular, convolutional neural networks (CNNs) have been very successful at classifying input images into a set of output classes. This has been demonstrated in the recognition of hand-written digits~\cite{lecun98gradient}, in the classification of Web images~\cite{krizhevsky12imagenet} and in the annotation of online videos~\cite{karpathy14video}, to cite only a few applications.

In general, CNNs have a common overall structure, which is depicted in Figure~\ref{fig:cnn}. This comprises, namely:
\begin{itemize}
	
\item An input layer, which receives an image, a multi-channel image (in case of color images), or a set of video frames.

\item One or more convolutional layers, where multiple filters (in the form of a sliding window) are applied to the same input. Each filter produces a different \emph{feature map}.

\item A subsampling layer after each convolution or, alternatively, some form of subsampling applied during the convolution itself, namely by making the sliding window \emph{stride} in larger steps across the input.

\item One or more densely-connected layers at the end to perform classification based on the features extracted by the convolutional layers.

\item An output layer with the same number of nodes as the number of output classes.

\end{itemize}

The idea is to have a first stage of convolutional layers to extract meaningful features from the input image, and a second stage of dense layers to perform the actual classification based on those features. Every time a convolutional filter is applied, it produces a new feature map as output. The first convolutional layer operates directly on the input image; subsequent layers operate on the feature maps produced by the previous convolutional layer. The purpose of having subsampling (or a stride greater than 1) is to make the feature maps smaller and allow their number to progressively increase, while keeping the network under a manageable size.

In a typical CNN, the input is a 2D image and the output is a 1D vector of class probabilities. In tomography, however, the scenario is the opposite: the input is a 1D vector of bolometer measurements and the output is a 2D image of the plasma radiation profile. Therefore, for the purpose of tomographic reconstruction, it makes sense to think of the inverse of a CNN, i.e.~a kind of ``deconvolutional'' network that is able to reconstruct a 2D image from its 1D projections.

In the literature, deconvolutional neural networks have been used for image generation~\cite{dosovitskiy15learning,dosovitskiy17learning}. By specifying the class label and the camera position (1D data), the network is able to generate an object (2D image) of the specified class from the given camera view. In this case, the network architecture is basically the reverse of a CNN, with a couple of dense layers at the beginning, and a series of convolutional layers (with upsampling) to generate the output image.

In previous work~\cite{matos17deep}, we have shown that such kind of network can produce tomographic reconstructions with high accuracy when trained on certain subsets of JET pulses. In the present work, we improved the network architecture by replacing the convolution+upsampling layers with proper deconvolutional layers (i.e.~transposed convolutions~\cite{dumoulin16guide}) in order to obtain the logical inverse of a CNN. We have also removed the requirement for any data preprocessing, so the measurement data coming from the bolometers can be fed directly to the network. These improvements allow the network to be trained much faster and with larger amounts of data.

\begin{figure*}[t]
	\centering
	\includegraphics[width=\textwidth]{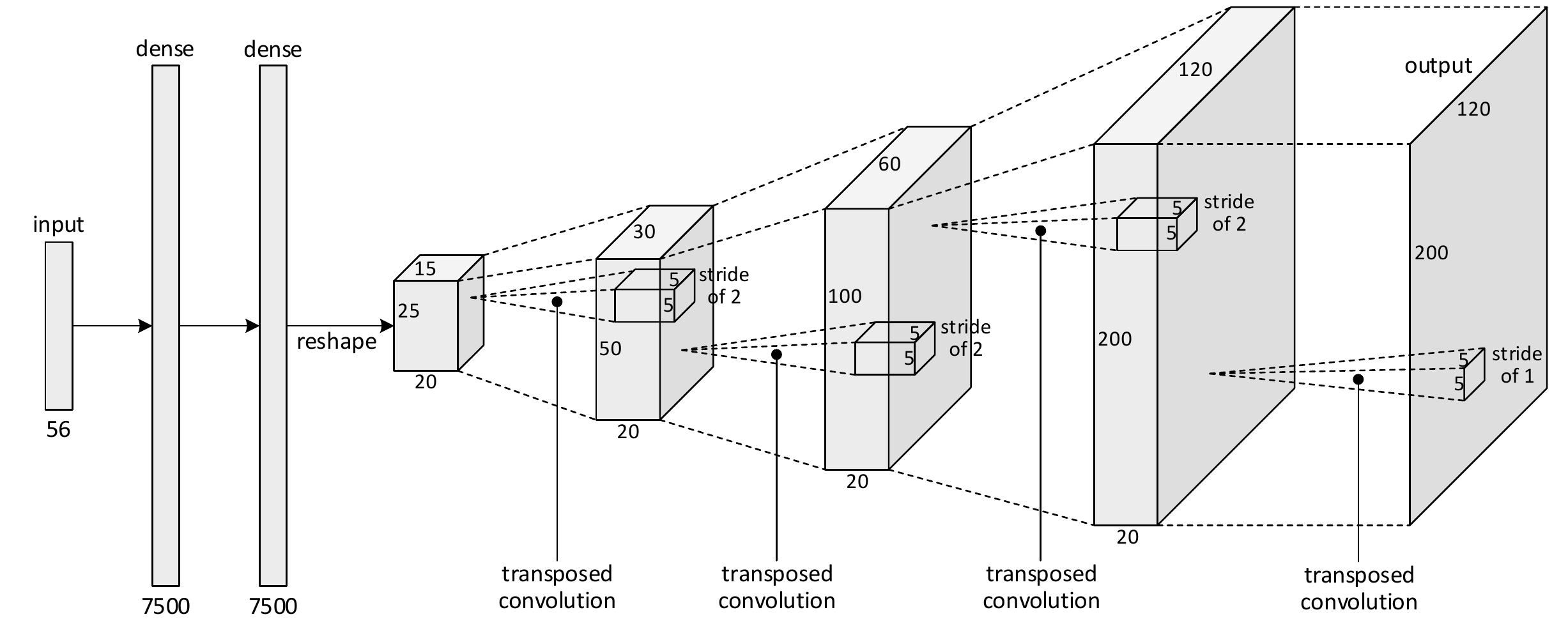}
	\caption{Deconvolutional neural network for tomographic reconstruction}
	\label{fig:deconv}
\end{figure*}

The resulting network architecture is depicted in Figure~\ref{fig:deconv}. After the input layer for the 56 bolometer channels, there are two dense layers with 7500 nodes, which can be reshaped into a 3D structure of size 25$\times$15$\times$20. This structure can be regarded as comprising 20 features maps of size 25$\times$15. By applying a series of transposed convolutions, these features maps can be brought up to a size of 200$\times$120, from which the output image is generated by one last convolution.

The transposed convolution is the inverse of the convolution in the sense that, if a sliding window would be applied to the output, the result would be the feature map given as input. It can be shown that learning a transposed convolution is equivalent to learning a weight matrix that is the transpose of a regular convolution~\cite{dumoulin16guide}, hence the name of this operation.

In Figure~\ref{fig:deconv}, the upsampling of the feature maps from 25$\times$15 to 200$\times$120 is achieved by having each transposed convolution operate with a stride of two pixels (i.e.~one pixel is being skipped between each two consecutive positions of the sliding window). This means that the output is two times larger and taller than the input, except for the very last convolution which uses a stride of one to keep the same size.

\section{Training and accuracy}

In order to train the network in Figure~\ref{fig:deconv}, we gathered as many sample reconstructions as possible. For this purpose, we collected every single reconstruction that has been produced at JET since the installation of the ITER-like wall (ILW) in 2011~\cite{matthews11ilw}. This yielded a total of 24203 sample tomograms, which have been separated into 90\% (21783) for training and 10\% (2420) for validation.

The network was trained using an adaptive gradient descent algorithm~\cite{kingma14adam} with a relatively small learning rate ($10^{-4}$) and a large batch size (411). The batch size was chosen in order to have about 50 batches, and in order to avoid having any partially filled batch in the training set. Since 411 is a divisor of 21783, this gives 21783/411=53 batches, and therefore there are 53 updates per epoch, where an epoch is one run through the whole training set.

\begin{figure}[h]
	\centering
	\includegraphics[width=\linewidth]{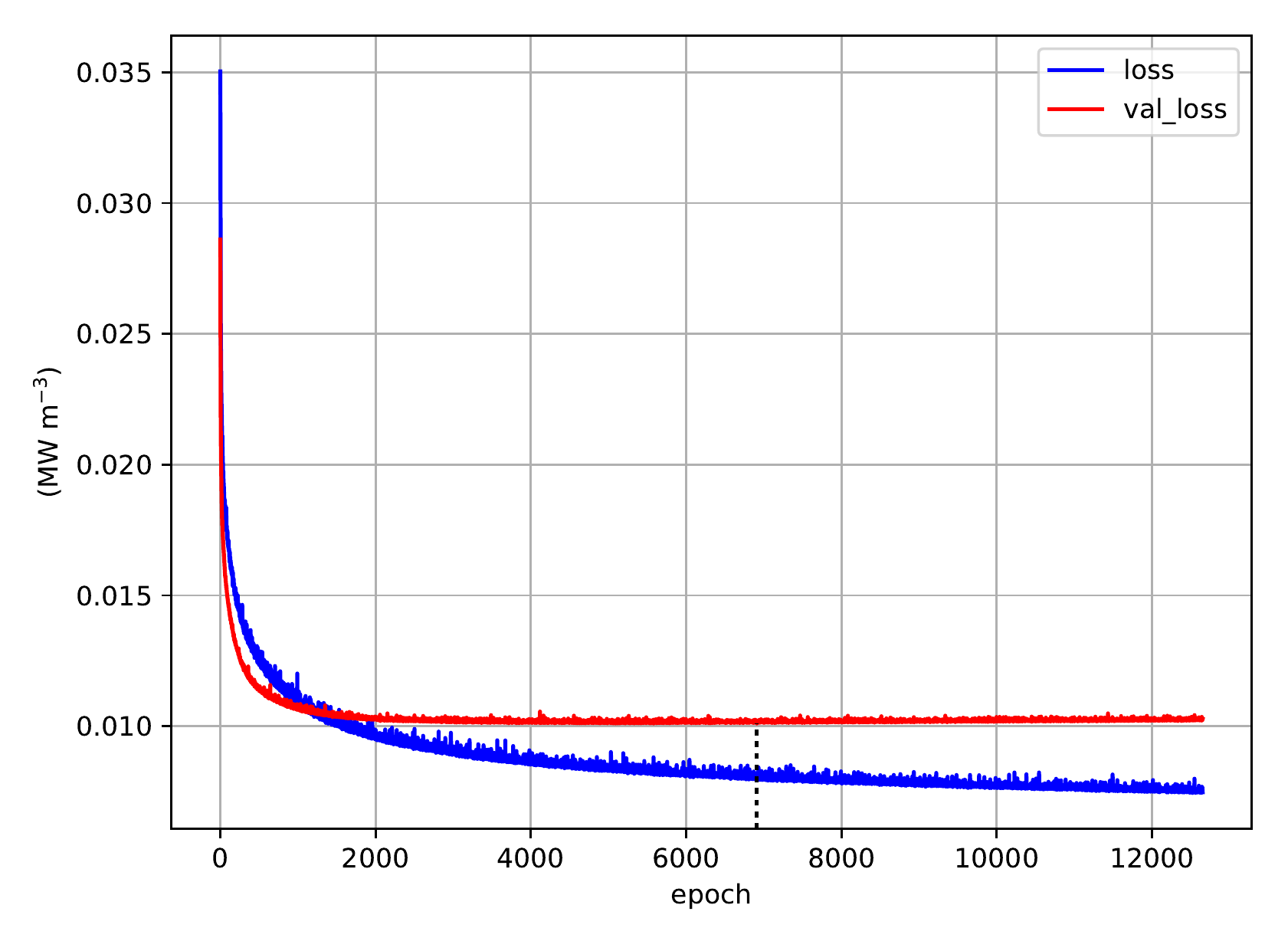}
	\caption{Loss and validation loss during training}
	\label{fig:plot_train}
\end{figure}

The network was trained on an Nvidia Titan X graphics processing unit (GPU) until the error in the validation set no longer improved. After 60 hours (12652 epochs) the minimum error on the validation set was achieved at epoch 6911, and there was no improvement after that. In fact, after that point the error on the validation set started to increase (this is marginally visible in Figure~\ref{fig:plot_train}), which should be taken as a symptom of overfitting. Therefore, we kept the network weights from epoch 6911.

The minimum error on the validation set, measured as (pixel-wise) mean absolute error, was about 10~kW~m$^{-3}$. The relatively small size of this error can be appreciated by comparing it to the dynamic range of the sample reconstruction in Figure~\ref{fig:lines_tomo}.

\begin{figure}[t]
	\centering
	\begin{tabular}{c}
		\includegraphics[width=0.95\linewidth]{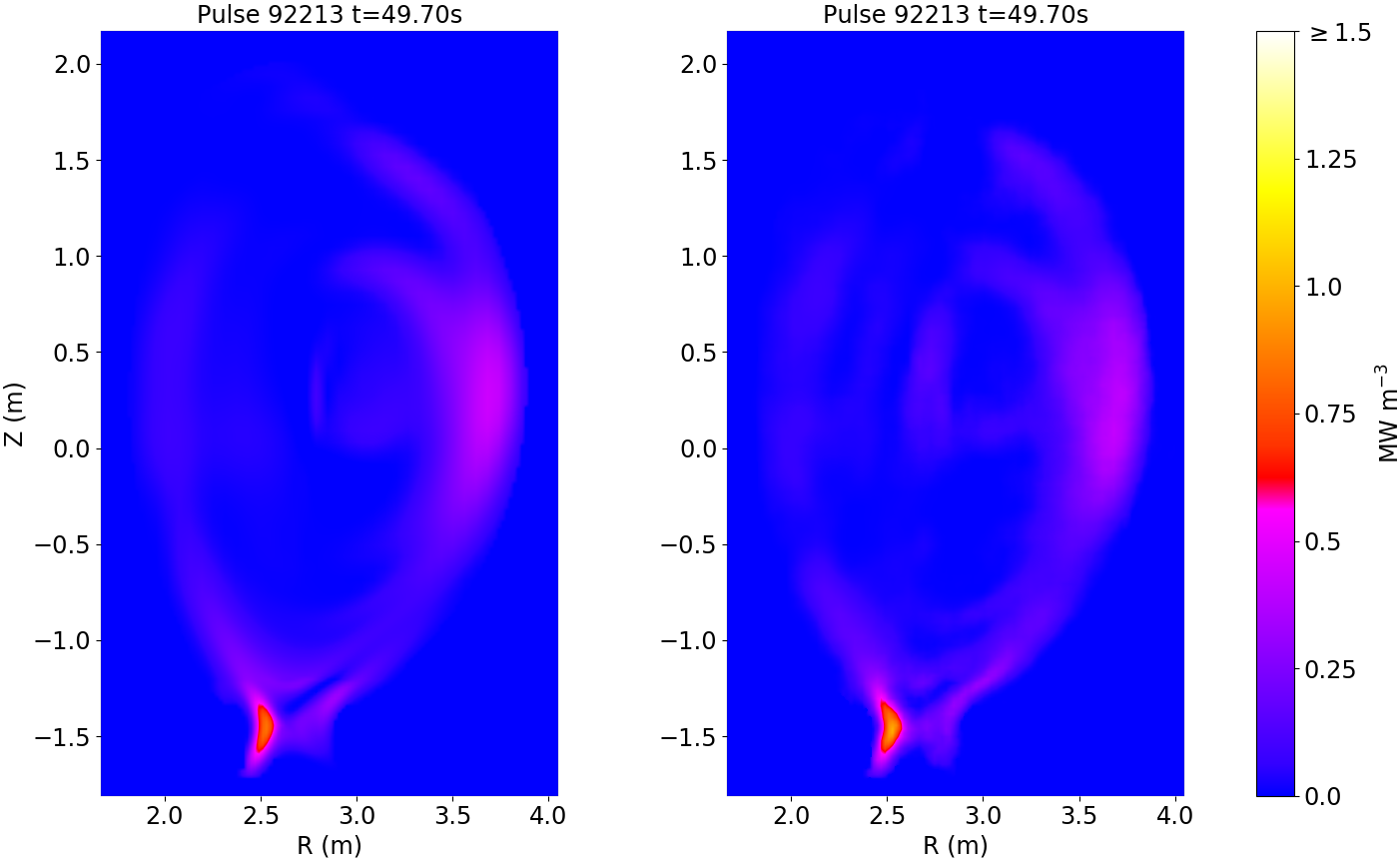} \\
		\includegraphics[width=0.95\linewidth]{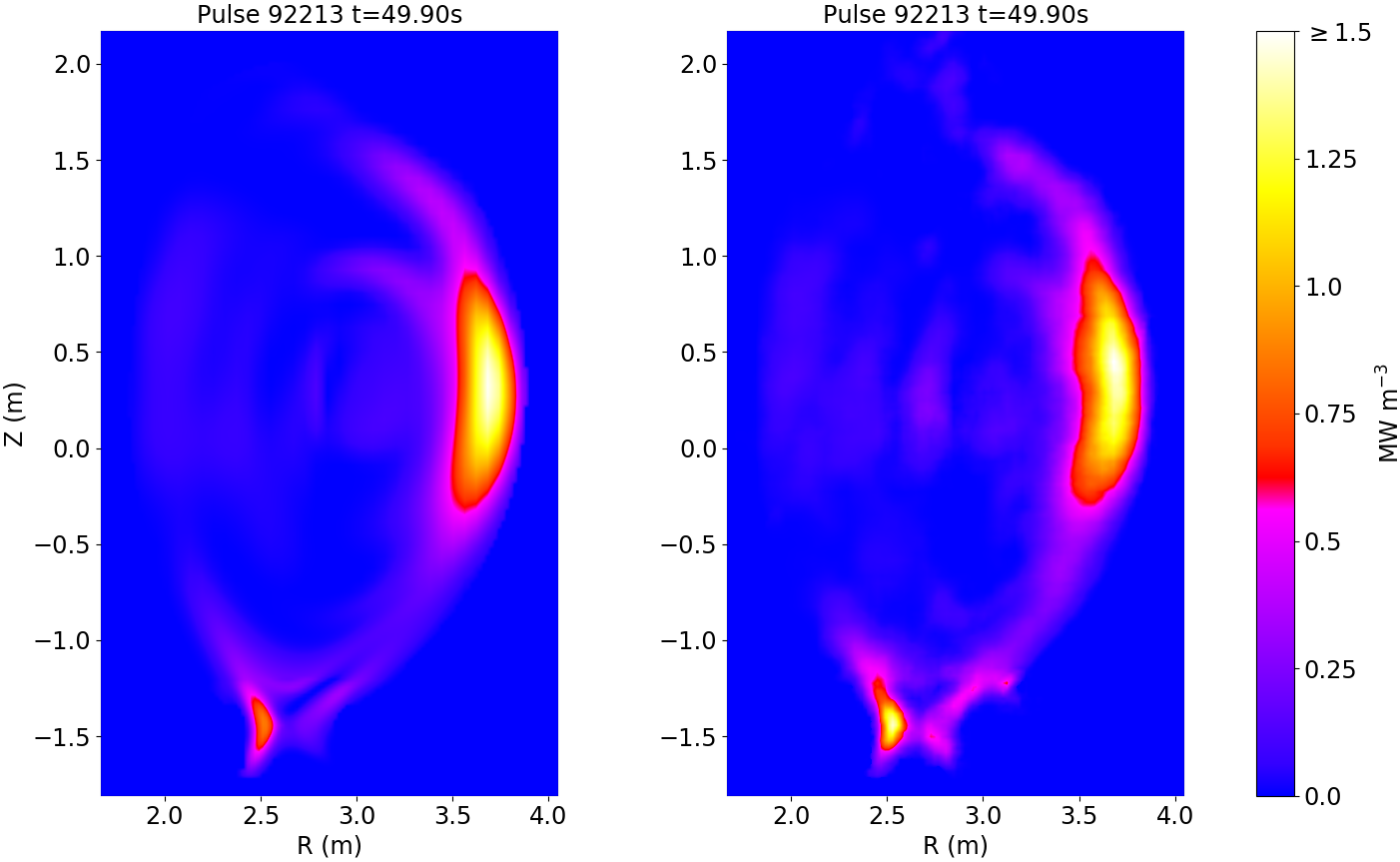} \\
		\includegraphics[width=0.95\linewidth]{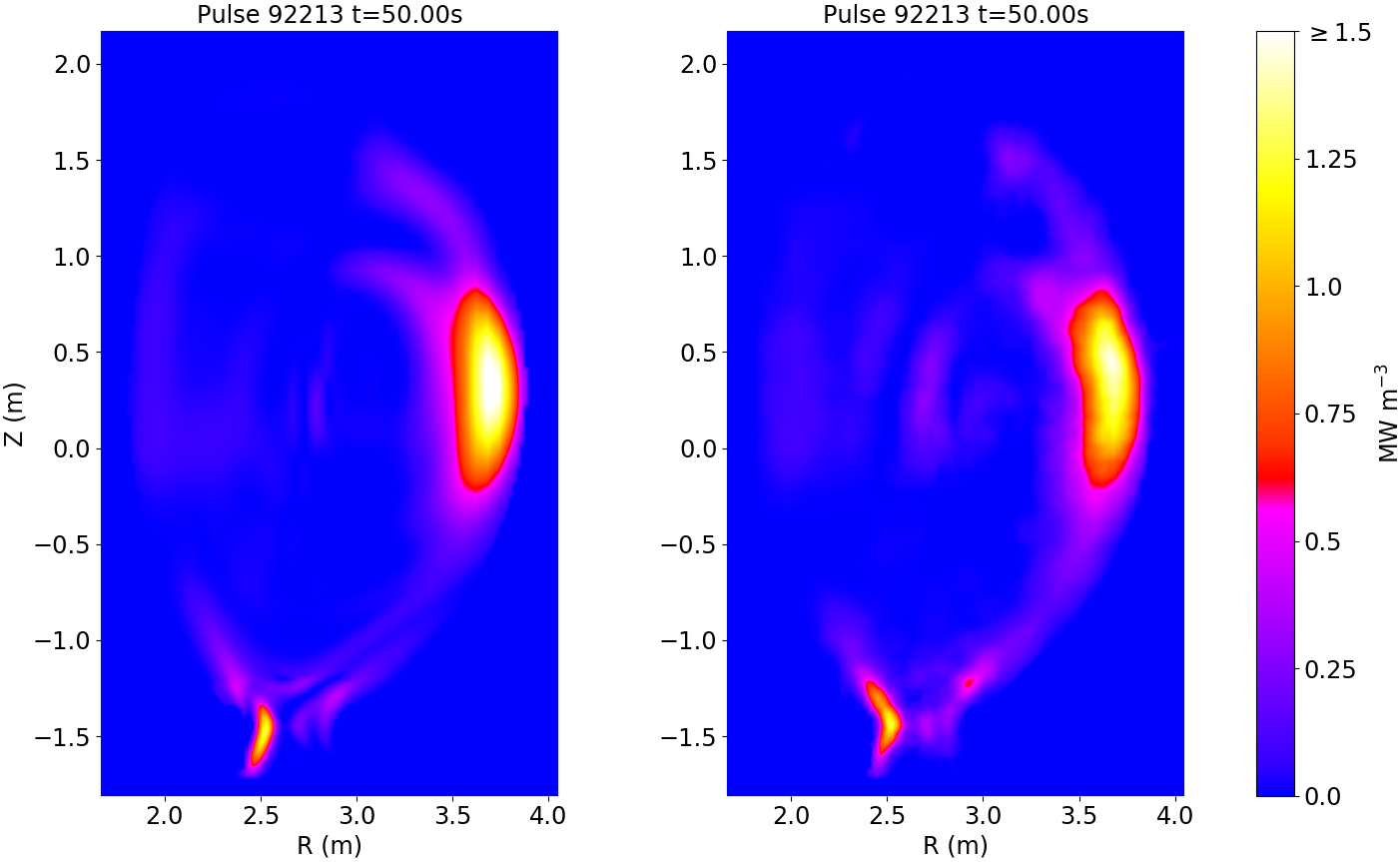}
	\end{tabular}
	\caption{Three reconstructions from pulse 92213 at t=49.7s (\emph{top}), t=49.9s (\emph{middle}) and t=50.0s (\emph{bottom}). Original reconstruction (\emph{left}) vs. neural network (\emph{right}).}
	\label{fig:compare}
\end{figure}

As an example of the results that can be obtained with the trained network, Figure~\ref{fig:compare} shows three reconstructions from pulse 92213. On the left is the reconstruction produced by the iterative constrained optimization method, and on the right is the reconstruction produced by the neural network. It should be noted that these particular reconstructions were not part of the training set or the validation set.

Despite some perceptible differences, it is clear that the neural network knows exactly where the focus of radiation is (top and middle row) and can adapt even to small changes in shape (middle and bottom row). To quantify the differences between each pair of reconstructions, it is possible to use image quality metrics such as structural similarity (SSIM)~\cite{wang04ssim}, peak signal-to-noise ratio (PSNR)~\cite{thu08psnr} and normalized root mean square error (NRMSE)~\cite{fienup97nrmse}. Table~\ref{tab:metrics} shows the kind of results that can be expected with these metrics.

\begin{table}[t]
	\centering
	\small
	\begin{tabular}{lrrr}
		\toprule
		time & SSIM & PSNR & NRMSE \\
		\midrule
		t=49.7s & 0.9354 & 30.386 & 0.06925 \\ 
		t=49.9s & 0.9239 & 28.161 & 0.08437 \\
		t=50.0s & 0.9134 & 28.864 & 0.07750 \\
		\bottomrule
	\end{tabular}
	\caption{Quality metrics on the reconstructions of Fig.~\ref{fig:compare}}
	\label{tab:metrics}
\end{table}

\begin{table}[t]
	\centering
	\small
	\begin{tabular}{lrrr}
		\toprule
		& SSIM & PSNR & NRMSE \\
		\midrule
		mean & 0.9573 & 33.791 & 0.0678 \\ 
		std. dev. & 0.0609 & 6.064 & 0.0415 \\
		\bottomrule
	\end{tabular}
	\caption{Quality metrics on the validation set}
	\label{tab:validate}
\end{table}

The examples in Figure~\ref{fig:compare} have not been carefully selected to yield particularly good results and, in practice, it might be possible to obtain even better results. For example, Table~\ref{tab:validate} shows the results obtained on the validation set, where all metrics are, on average, slightly better. Also, in previous work~\cite{matos17deep} we obtained even better results but it should be noted that, in that case, the network was trained separately on smaller and more uniform subsets of data.

In this work, the network was trained on a wide range of pulses, from 80000 (Aug.~2011, after the installation of the ILW) to 92504 (Nov.~2016, the last pulse at the time of this writing). It is possible that, during these 5 years, the bolometer system suffered some degradation (some channels might be broken or malfunctioning occasionally) and yet the neural network can make sense of the bolometer data to provide accurate reconstructions across this wide range of pulses.

\section{Full-pulse reconstruction}

Once trained, the network can be used to generate the reconstruction for any given pulse and time. In fact, on the same GPU that has been used before, the trained network can produce about 3000 reconstructions per second (or 100 reconstructions per second on a standard quad-core CPU). This means that the computation time for 6000 reconstructions (which would take 250 days, as mentioned in the introduction) can be brought down to just 2s, making it possible to generate at once all the reconstructions for an entire pulse.

\begin{figure*}[t]
	\centering
	\includegraphics[width=\textwidth]{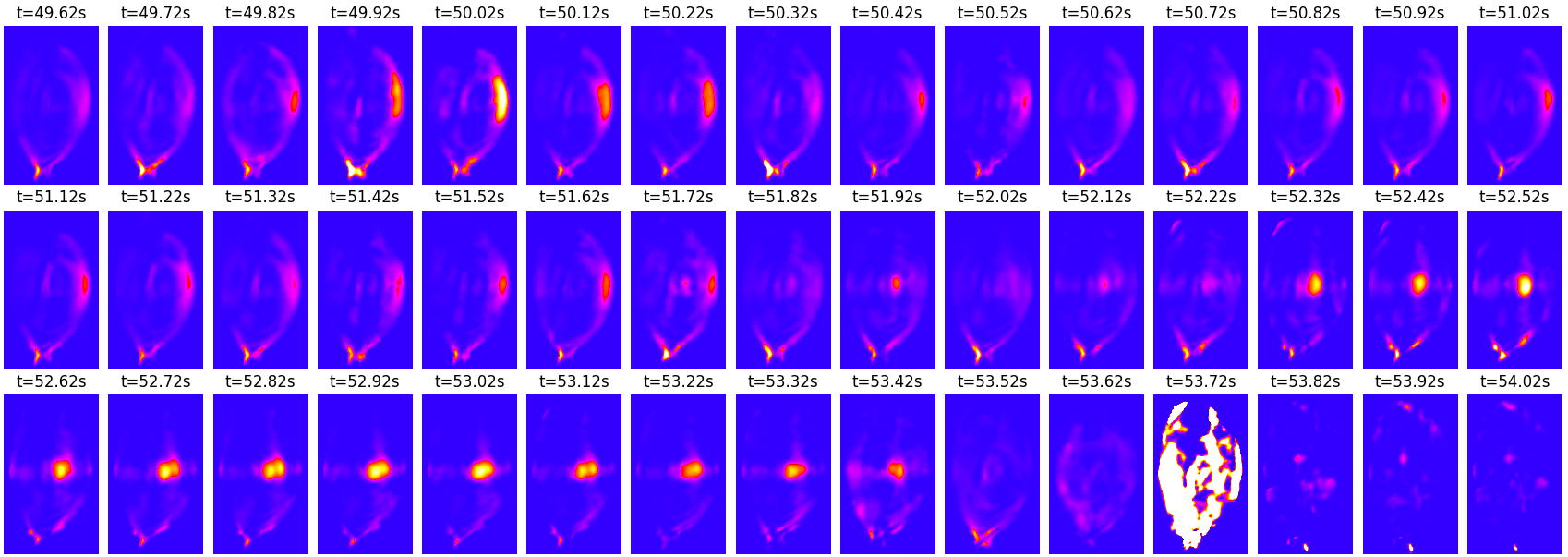}
	\caption{Reconstruction of pulse 92213 from t=49.62s to t=54.02s with a time step of 0.1s}
	\label{fig:92213}
\end{figure*}

Figure~\ref{fig:92213} shows the reconstruction of pulse 92213 from t=49.62s onwards, with a time increment of 0.1s (due only to space restrictions). The first row shows a focus of radiation developing on the outer wall, as we have seen in Figure~\ref{fig:compare}. However, Figure~\ref{fig:92213} also shows what happens afterwards: the focus of radiation seems to slowly fade away (rows 1, 2), only to reappear later at the core with particularly strong intensity (rows 2, 3). (The dynamic range of these frames is the same as in Figure~\ref{fig:compare}, i.e. 0 to 1.5 MW m$^{-3}$.)

\begin{figure*}[b]
	\centering
	\includegraphics[width=\textwidth]{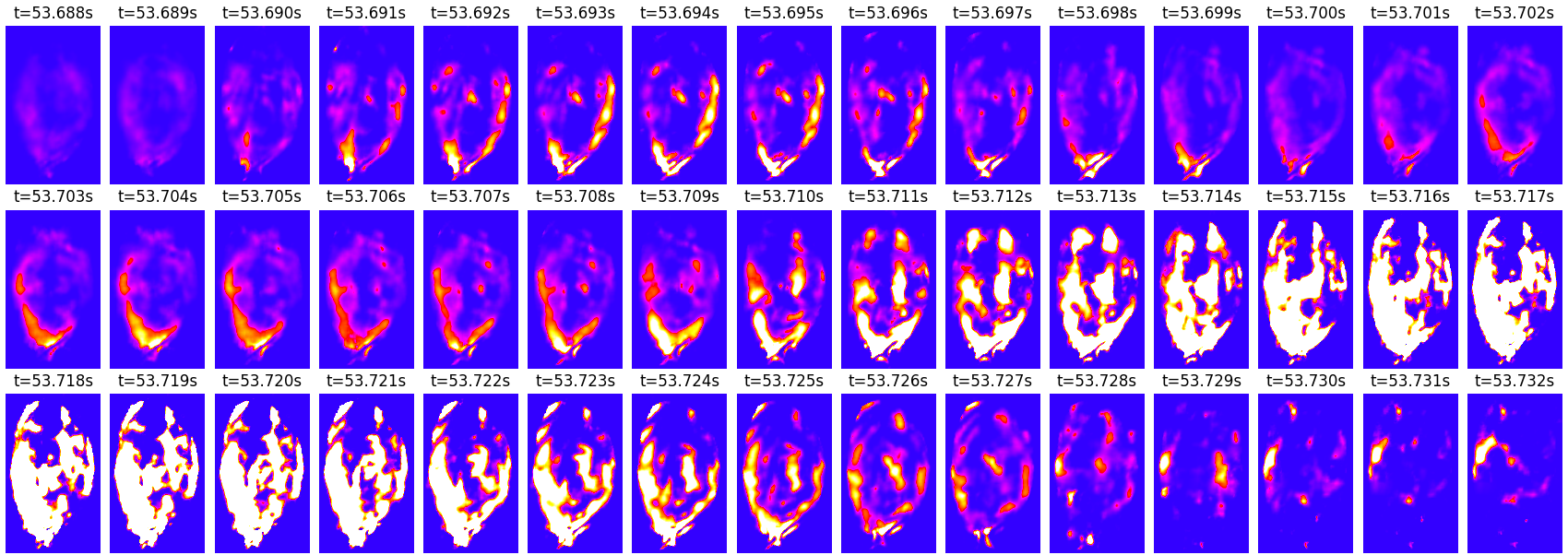}
	\caption{Reconstruction of pulse 92213 around disruption time with a step of 0.001s}
	\label{fig:disruption}
\end{figure*}

The radiation peaking stays in the core for a relatively long time (at least from t=52.32s to t=53.42s), while changing slightly in shape during that interval. Eventually, it also fades away as the heating systems are being turned off. However, just as it seemed that the plasma was about to soft land, there is a disruption at t=53.72s.

Clearly, there are phenomena of impurity sputtering, transport and accumulation in this pulse. Heavy ions (e.g. tungsten) coming from the wall and the divertor can lead to significant radiation losses, and these phenomena are especially noticeable in the radiation profile. It is therefore not surprising that bolometer systems are one of the key diagnostics used in disruption studies~\cite{arnoux09heat,riccardo10disruption,huber11radiation,lehnen11disruption}, either by providing a measurement of total radiation across the pulse, or by providing the tomographic reconstruction at specific points of interest, e.g.~during the thermal or current quench.

With full-pulse reconstruction, it is possible to carry out such analysis in more detail. For example, we can zoom into the final moments of pulse 92213 to observe what happens at the onset of (and even during) the disruption. Figure~\ref{fig:disruption} shows the reconstruction from t=53.688s to t=53.732s with a time increment of just 1ms. In these frames, there is an apparent swing from the outer wall to the inner wall, before the divertor lightens and a full-blown disruption occurs.

Although the dynamics of disruptions are not yet fully understood, one of the main causes is impurity accumulation in the core, which decreases the core temperature, creating a hollow temperature profile that eventually leads to core collapse~\cite{vries11survey,vries12impact}. There are a number of experiments at JET where impurities are deliberately injected into the plasma in order to study impurity transport and how impurities get to the core~\cite{pasini90impurity,mattioli95laser,galli98injection,solano16radiation}. In these experiments, the impurities are created by laser ablation of metal plate, and are subsequently injected into the plasma from the outer wall.

\begin{figure*}[t]
	\centering
	\includegraphics[width=0.985\textwidth]{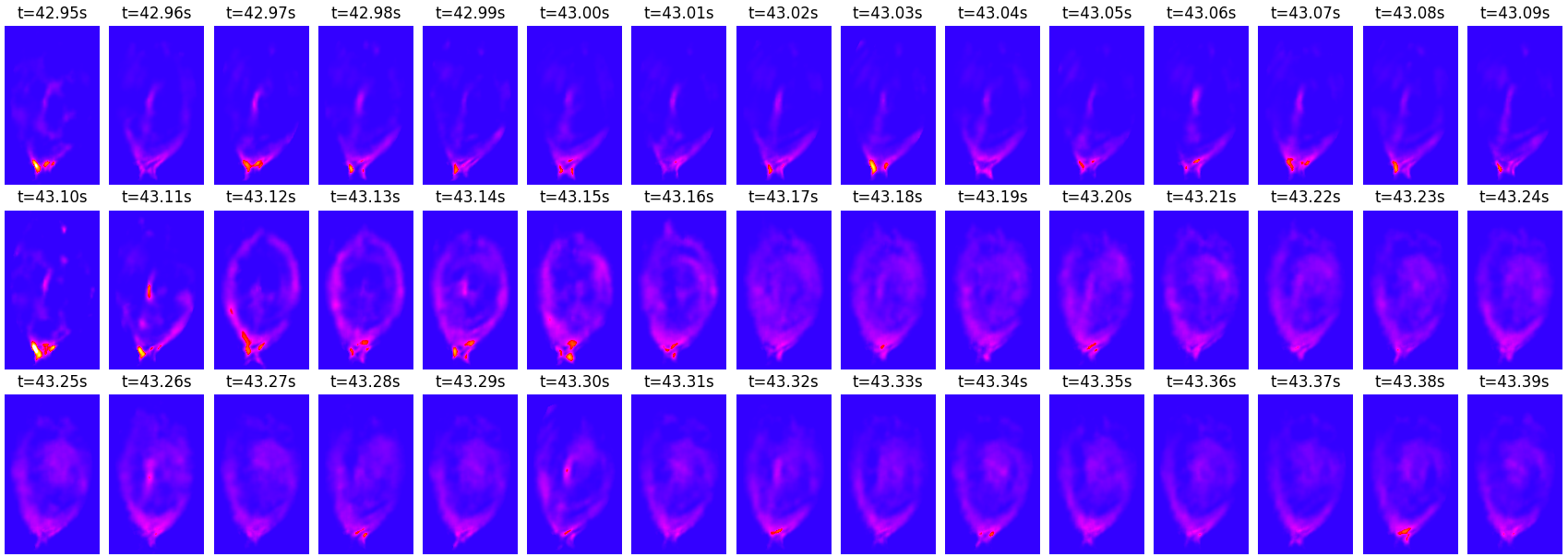}
	\caption{Reconstruction of pulse 92286 from t=42.95s to t=43.39s with a time step of 0.01s}
	\label{fig:92286}
\end{figure*}

Figure~\ref{fig:92286} shows an excerpt of the full-pulse reconstruction for an impurity-injection experiment (92286). Initially, the pulse is relatively quiet, but in the second row there is a noticeable change, with the formation of a ring around the edge. This is the moment when impurities (in this case, tungsten) have been injected into the plasma (around t=43.1s). The ring turns into a cloud and, as the cloud dissipates, the impurities appear to have concentrated at the core.

\begin{figure*}[b]
	\centering
	\includegraphics[width=0.985\textwidth]{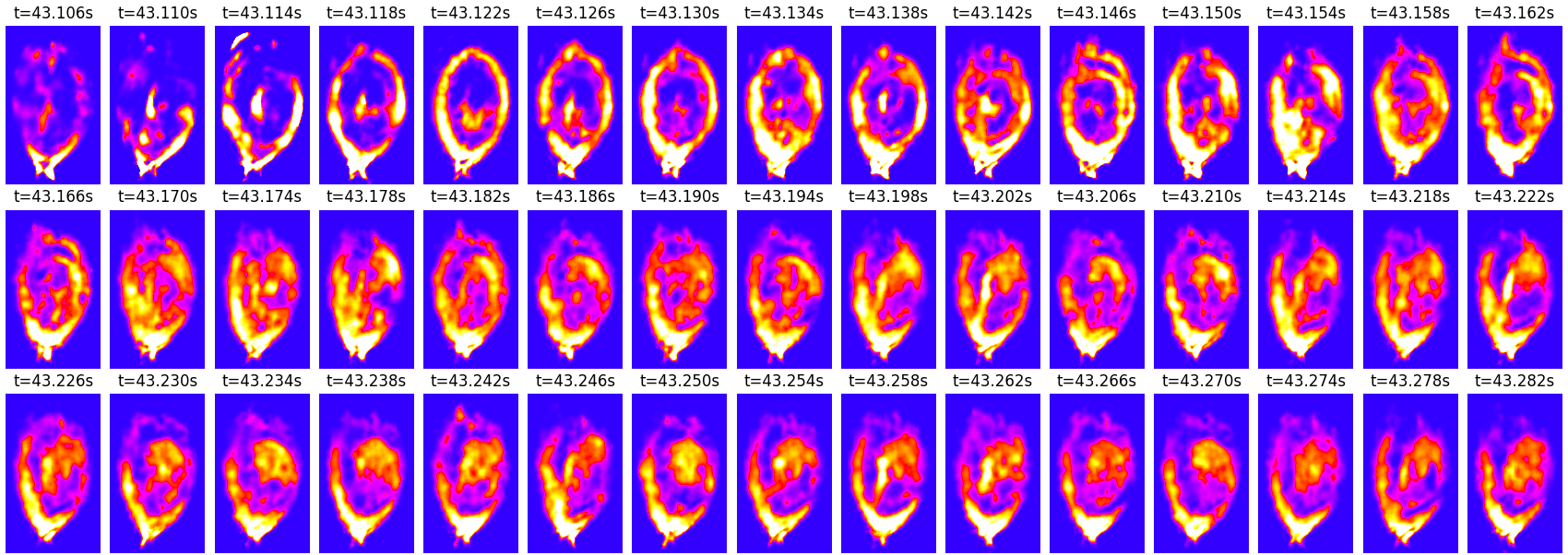}
	\caption{Reconstruction of 92286 pulse from t=43.106s to t=43.282s with a time step of 0.004s}
	\label{fig:ablation}
\end{figure*}

This impurity transport can be analyzed in more detail by decreasing the time step, and also by decreasing the dynamic range of these plots in order to have a better visualization. Figure~\ref{fig:ablation} shows a reconstruction with a time increment of 4ms and with a relatively low dynamic range of 0 to 30 kW m$^{-3}$. On the second frame in the top row (t=43.110s), the injection of tungsten is clearly visible at a midpoint on the outer wall.

Following the injection, the impurities flow in clockwise direction towards the divertor and  along the separatrix until forming a closed ring at t=43.122s. The ring thickens and the particles appear to follow a spiral movement towards the center. In the last row, a separate clump of radiation is already well established in the core, with another focus of activity going on in the divertor region.

\section{Conclusion}

A deep neural network, with an architecture that resembles the inverse of a traditional CNN, can be trained to produce tomographic reconstructions much faster than traditional methods. This makes it possible to reconstruct entire pulses, providing a detailed visualization of several phenomena, such as impurity transport and disruptions.

The same neural network approach can be applied to other diagnostics in the same machine, or to similar diagnostics in other machines, provided that there is sufficient training data. However, it should be noted that the neural network cannot produce better results than the method that was used to generate the training data itself. In general, the neural network is just an approximation with arbitrarily good precision and, as such, its results may suffer from the same problems or artifacts that are present in the training data.

The tomographic routines at JET have been stable and have been widely used for a number of years. This provided a consistent a relatively large training dataset. In other machines, such as ITER, it will not be possible to use neural network methods from the start of operations, for lack of training data. However, it may be possible to train a neural network based on the part of the physics that is understood and can be simulated, such as the large-scale simulations provided by magnetohydrodynamics and impurity transport codes.

As future work, we intend to use the full-pulse reconstructions to analyze the precursors of disruptions in order to be able to mitigate, or even avoid those disruptions, if at all possible. As a testbed for ITER, JET is perhaps the last opportunity to tackle this problem before having to deal with it on a much larger scale.

\section*{Acknowledgments}

This work has been carried out within the framework of the EUROfusion Consortium and has received funding from the Euratom research and training programme 2014-2018 under grant agreement No 633053.
%The views and opinions expressed herein do not necessarily reflect those of the European Commission.
IPFN activities also received financial support from \emph{Funda\c{c}\~{a}o para a Ci\^{e}ncia e a Tecnologia} (FCT) through project UID/FIS/50010/2013. The Titan X GPU used in this work was donated by NVIDIA Corporation.

We would also like to thank Ewa Pawelec at the University of Opole, in Poland, for insights into laser ablation experiments, and Peter Lomas at JET, UKAEA, for several helpful discussions regarding disruptions.

\section*{Supplementary materials}

Videos of the reconstructed pulses presented in this paper (92213 and 92286) can be found online at: {\footnotesize\url{http://web.tecnico.ulisboa.pt/diogo.ferreira/videos/}}

\bibliographystyle{ans}
\bibliography{references}

\end{document}